\documentclass[journal=jpcafh,manuscript=article]{achemso}
\setkeys{acs}{articletitle = true}
\usepackage[version=3]{mhchem} 
\usepackage[T1]{fontenc}       
\usepackage{amssymb}
\usepackage{amsmath}

\usepackage{graphicx} 
\usepackage{color}

\usepackage{booktabs} 
\usepackage{array} 
\usepackage{paralist} 
\usepackage{verbatim} 
\usepackage{subfig} 
\usepackage{hyperref}

\newcommand*{\citen}[1]{%
  \begingroup
    \romannumeral-`\x 
    \setcitestyle{numbers}%
    \cite{#1}%
  \endgroup   
}

\makeatletter
\newcommand*{\rom}[1]{\expandafter\@slowromancap\romannumeral #1@}
\makeatother

\title{Nonadiabatic quantum dynamics of tribovoltaic effects at sliding metal-semiconductor interfaces}

\author{Guangming Liu}
\affiliation{School of Science, Westlake University, Hangzhou, Zhejiang 310024, China }
\alsoaffiliation{Institute of Natural Sciences, Westlake Institute for Advanced Study, Hangzhou, Zhejiang 310024, China}

\author{Jun Liu}
\email{jliu238@buffalo.edu}
\affiliation{Department of Mechanical and Aerospace Engineering, University at Buffalo, The State University of New York, Buffalo, NY 14260, USA}
\alsoaffiliation{RENEW (Research and Education in Energy, Environment and Water) Institute, University at Buffalo, The State University of New York, Buffalo, NY 14260, USA}
\author{Wenjie Dou}
\affiliation{School of Science, Westlake University, Hangzhou, Zhejiang 310024, China }
\alsoaffiliation{Institute of Natural Sciences, Westlake Institute for Advanced Study, Hangzhou, Zhejiang 310024, China}
\email{douwenjie@westlake.edu.cn}
\SectionNumbersOn

\begin{document}

\begin{abstract}

Recent experiments observe electric current generation at a sliding metal-semiconductor interfaces. Here, we present a detailed theoretical study on how electric voltage is generated at such a sliding interface. 
Our study is based on a two-band Anderson-Holstein model, and we solve the coupled electron-phonon dynamics using a surface hopping method.  
We show that the high local temperature induced by mechanic motion at the interfaces could lead to electron-hole pair generation through electron-phonon couplings. 
We quantify the efficiency of electron-hole generation as well as electric voltage as a function of local temperatures and semiconductor bandgaps. We find that increasing the local temperatures can lead to higher electron-hole generations and electric voltage. Furthermore, we find that there is a turnover for the electric voltage as a function of the bandgap. Such an observation is in agreement with the experimental results. Our study offers a theoretical framework to understand tribovoltaic effects from a quantum mechanical point of view, and our approach can be used to quantitively simulate realistic sliding metal-semiconductor junctions.

\end{abstract}

\maketitle

\section{Introduction}
When a semiconductor material is involved in a sliding contact, a high direct-current (DC) can be generated, 
which has been recently observed in various contact material system (e.g. metal/semiconductor,\cite{liu2018direct,liu2019scaled,liu2018sustained} metal/insulator/semiconductor,
\cite{liu2018sustained,hao2019co} p-n junction,\cite{xu2019direct,zheng2020scanning}liquid/semicond-
uctor\cite{lin2020tribovoltaic,lu2020interfacial,zheng2021photovoltaic}). Such a phenomenon is referred to as 'tribovoltaic' effect\cite{zhang2020tribovoltaic}.
Moreover, new multi-physics phenomenon originated from the tribovoltaic effect such as tribo-photovoltaic effect\cite{liu2019separation,hao2019co,zheng2021photovoltaic} and tribovoltaic-thermoelectric effect
\cite{zhang2021tribo} have been found successively in dynamic metal-semiconductor Schottky systems.
Although the fundamental mechanism remains unsolved, it is found the DC current density output of such systems is on the order of 10-100 $A/m^2$, which is 3-4 orders’ 
higher than that in traditional piezoelectric or triboelectric nanogenerator and thus show a great promise on next-generation mechanical energy harvesting devices for self-powered electronics and 
Internet of Things (IoTs) sensors\cite{yang2021semiconductor}. 

In the past decades, tremendous efforts have been devoted to understanding and quantifying charge transfer during contact. To date, the fundamental physics of triboelectricity remains unsolved\cite{lacks2019long,lowell1980contact,wang2019origin}. 
The challenge of an explicit understanding of the mechanism resides in the complex multi-scale and multi-physics interaction at a contact interface. \cite{alicki2020quantum,willatzen2020quantum,antony2021electronic,abdelaziz2018atomistic} 
Frictional energy dissipation in materials may manifest in different ways: heat generation, exo-electron, luminescence, electron-hole generation, surface charges, etc\cite{park2014fundamental}. 
In general situation, the majority of frictional energy ends up with heat either through phonon-phonon interaction or relaxation of other electronic excitations. 
When a Schottky contact consisting of a metal and a semiconductor material is presented however, it is proposed that the input frictional energy may induce a non-adiabatic electronic excitation of energetic carriers, 
and the generated electron-hole pairs may be 
non-adiabatic electronic excitation, which ultimately contribute to the DC generation at a sliding Schottky contact\cite{liu2019interfacial,zhang2020tribovoltaic}.

The physical essence of charge transfer at the interfaces is the breakdown of Born-Oppenheimer (BO) approximation: 
In the BO approximation, nuclear motion is assumed to be much slower than electronic motion, such that nuclei move on the adiabatic potential energy surface. 
Under external stimuli, the accelerated nuclear motion can introduce non-adiabatic electronic transitions between different potential energy surface near the crossing point such that the BO approximation breaks\cite{dou2020nonadiabatic,nienhaus2002electronic}.  
Nevertheless, a basic understanding of the electronic excitation process in an electro-mechanical coupling system is still missing. 
In this work, a systematic study on the non-adiabatic electronic excitation at a metal/semiconductor sliding interface using quantum dynamics approach is presented, in an effort to shed light on fundamental process and provide instructional guide for materials design and optimization for future semiconductor-based nanogenerators.

In our previous studies, we have presented a surface hopping dynamics to describe nonadiabatic dynamics near a metal surface. \cite{dou2015surface,dou2015surface3} In such a dynamical method, we propagate trajectories on different potential energy surfaces describing different charge states. We further introduce stochastic hopping between them indicating charge transfer between these states. Note that, different Tully's surface hopping method, which is typically used to model nonadiabatic dynamics in solution or in gas phase, our surface hopping is intended to study nonadiabatic dynamics at metal surfaces. Our surface hopping dynamics have been successfully used to describe electron transfer, energy relaxation, current-voltage characteristic near metal surfaces as well as in nano-junctions. \cite{ouyang2016dynamics,dou2018broadened}

In this manuscript, we present a two-level Anderson-Holstein model based on electron-phonon couplings to describe the semiconductor-metal interfaces. 
We use a surface hopping method to describe the nonadiabatic dynamics at such interfaces. Our dynamics indicate that local temperature induced by mechanic motion at the interfaces could lead to electron-hole pair generation through electron-phonon couplings. We can then quantify the generation of the electron and hole population as well as the average voltage. We find that there is a turnover of the voltage as the function of the bandgap. 
This finding is in agreement with experimental results.  We believe that this article provides atomic insights on how electric current is generated on sliding semiconductor-metal interfaces. Our approach can be used to study realistic sliding metal-semiconductor junctions in the future.

\section{Theoretical model}   \label{sec:theory}
Here, we use a two-level/two-band Anderson-Holstein model to describe semiconductor/metal interfaces. 
Our total Hamiltonian consists of three parts: $\hat H_M$ for the metal, $\hat H_S$ for the semiconductor, and $\hat H_I$ for the couplings between the metal and semiconductors, i.e.
\begin{eqnarray}
\hat H = \hat H_M + \hat H_S + \hat H_I 
\end{eqnarray}
The metal consists of a manifold of electronic states, 
\begin{eqnarray}
\hat H_M = \sum_k \epsilon_k \hat c^+_k \hat c_k 
\end{eqnarray}
For simplicity, we model the semiconductor with two levels, conduction level/band $E_c  (x)$, and valence level/band $E_v (x)$. Here we have a local phonon degrees of freedom ($x$ and $p$ are the position and momentum of the phonon) coupled to the electronic levels. Such that the Hamiltonian for the semiconductor can be written as
\begin{eqnarray}
\hat H_S  = E_c  (x) \hat d_c^+ \hat d_c  + E_v (x) \hat d_v^+ \hat d_v   +  U_0 (x) + \frac{p^2}{2m}  
\end{eqnarray}
$U_0 (x)$ is the potential for the local phonon mode. Below, we will approximate the potential as a harmonic oscillator, $U_0= \frac12 m \omega^2 x^2$ ($\omega$ is the frequency of the oscillator). The couplings between the metal and levels in the semi-conductor are assumed to be bilinear: 
\begin{eqnarray}
\hat H_I = \sum_k V_{ck} ( \hat c^+_k d_c + d_c^+  \hat c_k  ) + \sum_k V_{vk} ( \hat c^+_k d_v + d_v^+  \hat c_k  )
\end{eqnarray}
We can define the hybridization functions to characterize the strength of the couplings:    
\begin{eqnarray}
\Gamma_c = \sum_k |V_{ck}|^2 \delta(\epsilon- \epsilon_k) \\
\Gamma_v = \sum_k |V_{vk}|^2 \delta(\epsilon- \epsilon_k) 
\end{eqnarray}
We further assume that the conductor band and the valence band couples to the local phonon as the following:
\begin{eqnarray}
E_c (x) = g x \sqrt{m\omega/\hbar}  + E_g/2 \\
E_v (x) = gx  \sqrt{m\omega/\hbar}   -  E_g/2 
\end{eqnarray}
Here, $g$ characterizes the electron-phonon coupling strength. We have assumed that the coupling is linearly proportional to local phonon displacement, which is the typical approximation for the electron-phonon interactions. $E_g$ is the band gap between the conduction band and valance band.

The above equations conclude our model. Our goal is to simulate the electron transfer processes between the conduction/valence band and metal surface. Such processes could lead to electron-hole pair generation in the semiconductor as well as electric voltage. Obviously, solving the real time dynamics will be very difficult. Below, we introduce a classical master equation and surface hopping (SH) method to simulate the dynamical processes.  

\begin{figure}[t] 
   \centering{}
   \subfloat[]{\includegraphics[width=3cm]{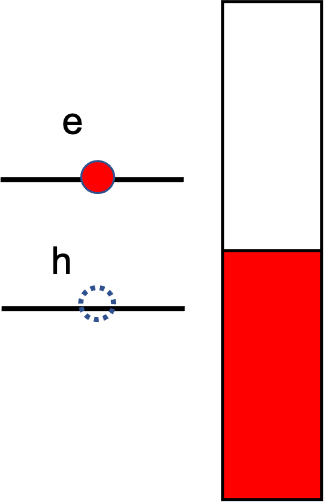}}
   \hfill
   \subfloat[]{\includegraphics[width=5cm]{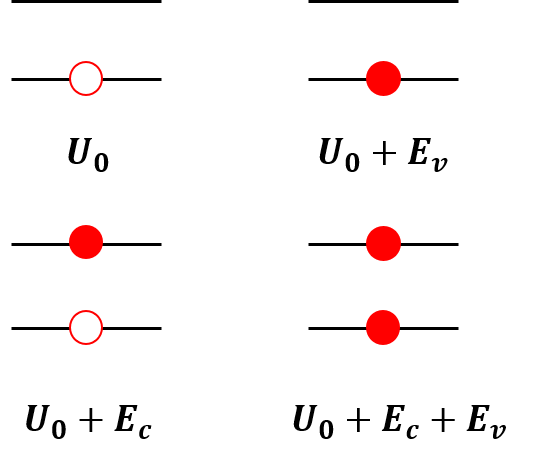}}
   \hfill
   \subfloat[]{\includegraphics[width=6cm]{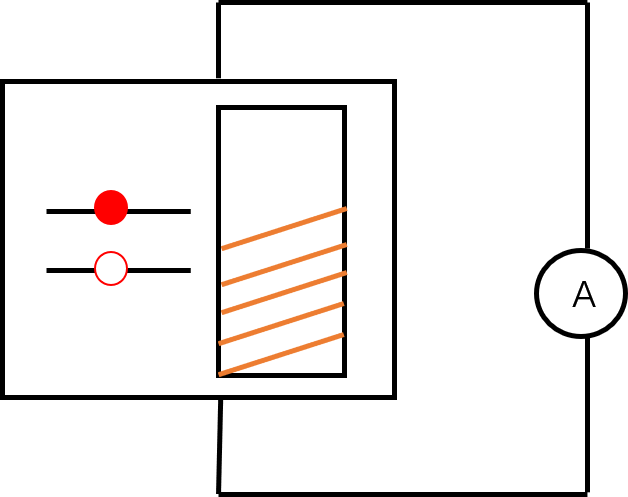}}
   \hfill
   \caption{Semiconductor near a metal surface. (a)Sliding semiconductor near a metal surface create electron and hole in the semiconductor. (b)Four states will arise after surface hopping. $U_0$: no electron in valance band(VB) or conduct band(CB), 
   $U_0+E_v$: one electron occupied in VB. $U_0+E_c$: one electron occupied in CB. $U_0+E_c+E_v$: two electrons occupied in CB and VB.
   (c)There will be electron current flow when connected to circus. Our model explain why electron-hole can be created at the surfaces. }
   \label{fig:mechanism}
\end{figure}

\subsection{Classical master equation} 

To solve the coupled electron-phonon motion at the interfaces for the two level Anderson-Holstein model introduced above, we explore the classical master equation method. 
In the limit of weak coupling ($\Gamma<kT$), provided phonon is classical ($\hbar\omega< kT$), we can derive a classical master equation (CME) to describe the dynamics. 
The exact derivation is similar to the one for one-level Anderson-Holstein model.\cite{dou2015surface,dou2015surface3} Here we outline the derivation and show the equations of motion.

 We start from the Liouville equation for the total system. Under the weak coupling assumption, after tracing over the bath degrees of freedom, we arrive at the quantum master equation for the system only: 
\begin{eqnarray}
\partial_t \hat \rho = i [ \hat H_s, \hat \rho] - \hat{\hat{\mathcal{L}}} \hat \rho 
\end{eqnarray}
Here, $\hat \rho$ is density operator for the system. $\hat{\hat{\mathcal{L}}} $ is the Lindblad superoperator coming from the system-bath coupling. Note that, the system includes the electronic states in the semiconductor as well as the phonon degrees of freedom. We further make classical assumption for the phonon using Wigner transformation: 
\begin{eqnarray}
\hat \rho (x, p) = \int d \Delta x \: \langle x-\Delta x | \hat \rho | x + \Delta x \rangle \exp(-i p\Delta x )
\end{eqnarray}
Here, $\hat \rho (x, p)$ can be interpreted as the phase space density for different electronic states. Particularly, we have $\rho_0$, $\rho_v$, $\rho_c$, $\rho_{vc}$ to denote the phase space densities with empty electron, one electron in the valance band, one electron in the conduction band, and two electrons, respectively. The equations of motion for these densities are:
\begin{eqnarray}
\partial_t \rho_0 = \frac{p}{m} \partial_x  \rho_0 -  \partial_x  U_0 \partial_p \rho_0 - (\Gamma_c f(E_c) + \Gamma_v f(E_v) )\rho_0 + \Gamma_c f(-E_c) \rho_c +  \Gamma_v f(-E_v) \rho_v \\
\partial_t \rho_v = \frac{p}{m} \partial_x  \rho_v -  \partial_x  U_1 \partial_p \rho_v +\Gamma_v f(E_v) \rho_0 - (\Gamma_c f(E_c) + \Gamma_v f(-E_v) )\rho_v + \Gamma_c f(-E_c) \rho_{vc}  \\
\partial_t \rho_c = \frac{p}{m} \partial_x  \rho_c -  \partial_x  U_2 \partial_p \rho_c + \Gamma_c f(E_c) \rho_0 - (\Gamma_c f(-E_c) + \Gamma_v f(E_v) )\rho_c + \Gamma_v f(-E_v) \rho_{vc} \\
\partial_t \rho_{vc} = \frac{p}{m} \partial_x  \rho_{cv} -  \partial_x  U_3 \partial_p \rho_{cv} + \Gamma_c f(E_c) \rho_c + \Gamma_v f(E_v) \rho_{v} -(\Gamma_c f(-E_c) + \Gamma_v f(-E_v) )\rho_{vc} 
\end{eqnarray}
Here, $U_1 = U_0+E_v$, $U_2= U_0 + E_c$, and $U_3 = U_0 + E_c + E_v$ correspond to the potential energy surfaces for one electron in the valance band, one electron in the conduction band, and two electrons, respectively. We note that, the densities follow the classical motion on the corresponding potential energy surfaces with transitions between different electronic states. The hopping rates are proportional to the hybridization function ($\Gamma_c$ or $\Gamma_v$) as well as the Fermi functions.  
In Figure~\ref{fig:mechanism}, we sketch a semiconductor with two levels near a continuum of electronic states (metal surface). For simplicity, we do not include electron-phonon couplings in our plots. 
Figure~\ref{fig:mechanism}(b) show the four electronic states with different electron occupations. We refer to the state with one electron on the valence band as our ground state, and the others as excited states. In particular,  the state with no electron is referred to as one hole state, and the state with one electron on the conduction band is referred to as electron-hole pair state.

\begin{figure}[t] 
   \centering{}
   \subfloat[]{\includegraphics[width=8cm]{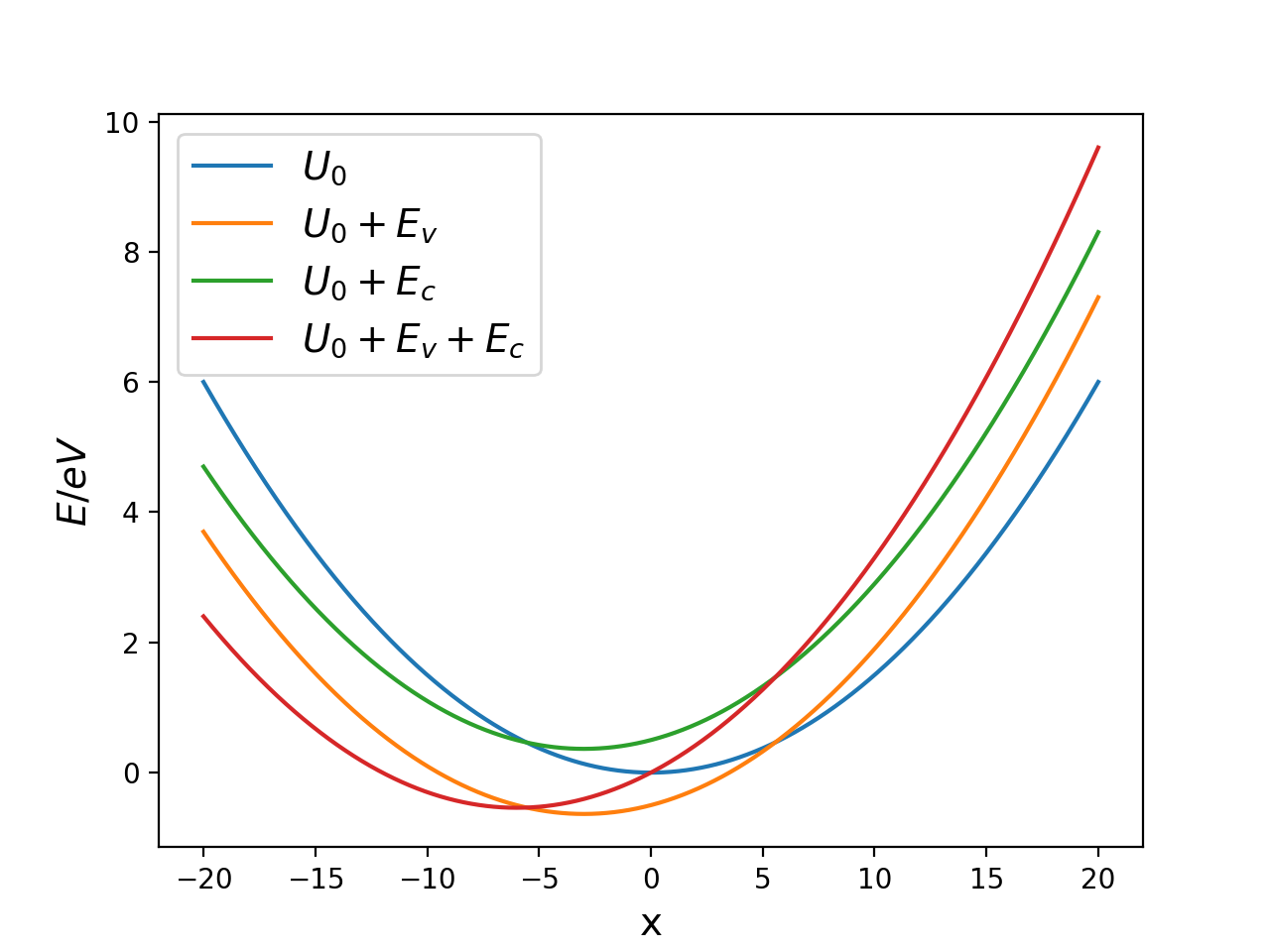}} 
   \hfill
   \subfloat[]{\includegraphics[width=8cm]{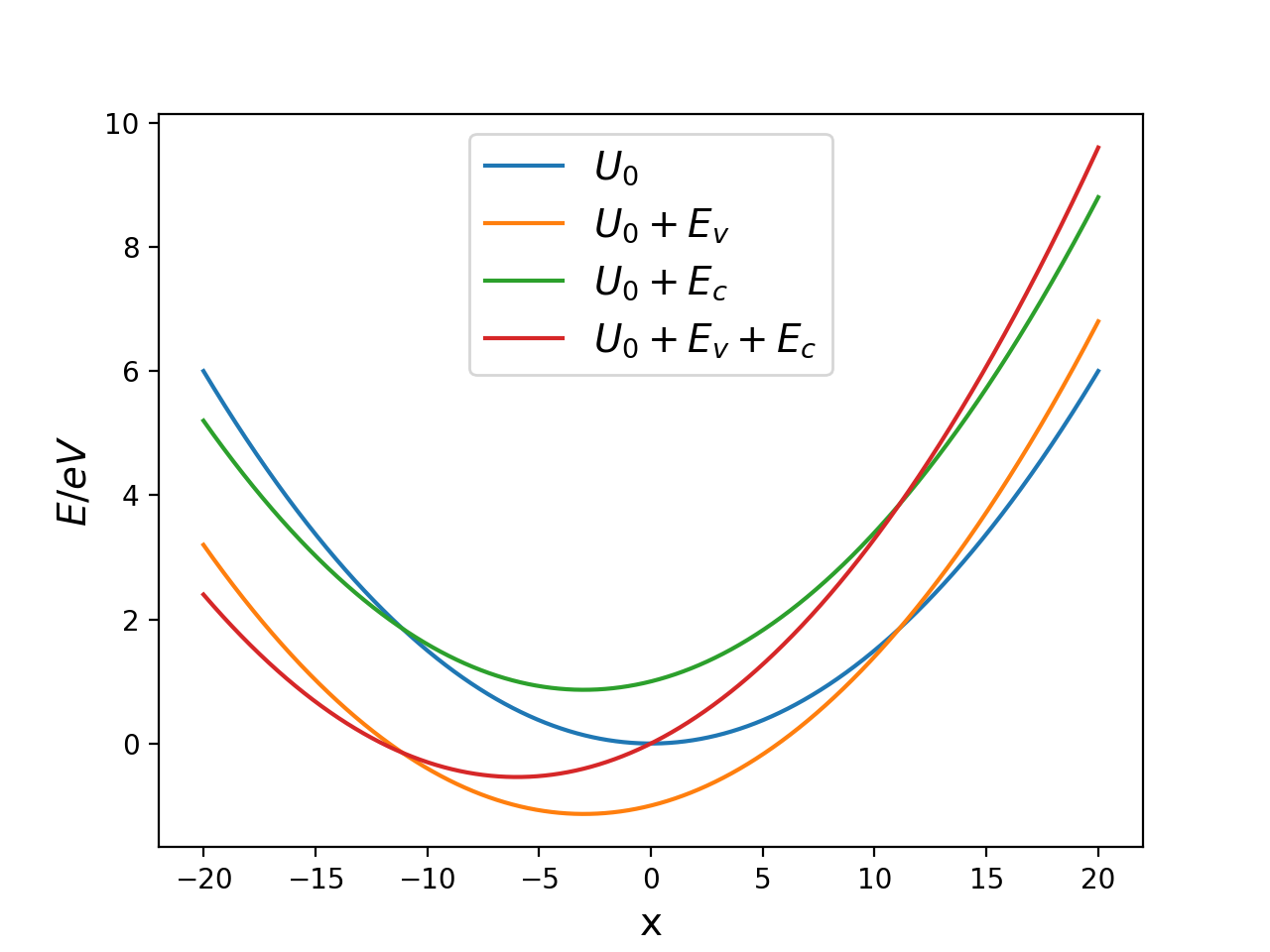}}
   \caption{potential energy surfaces for different charge states: no electron in either conduction or valance band $U_0$, one electron in the valence band $U_1= U_0+E_v$, 
   one electron in the conduction band $U_2= U_0 + E_c$, and two electrons in the conduction and valance band $U_3 = U_0 + E_v+ E_c$. Parameters: (a) $\hbar\omega=0.03\ eV$, $g_c = g_v = 0.09\ eV$, $E_g=1\ eV$ (b) $\hbar\omega =0.03\ eV$, $g_c = g_v = 0.09\ eV$,$E_g= 2\ eV$.}
   \label{fig:PES1}
\end{figure}

In Fig. \ref{fig:PES1}, we plot the potential energy surfaces for different electronic states. Notice that there are several crossings between the PESs when the bandgap $E_g$ is small. When the bandgap is large, the crossing can only happen at very high energies. At the crossing points, the hopping happens most frequently. At very low nuclear energy, the electrons will likely remain on the ground state. With higher nuclear energy, there will be more transitions on the excited states. Below, we will use a surface hopping algorithm to propagate the equation of motion and analyze non-adiabatic electron transfer between semiconductor and metal surfaces. 

\section{Numerical Results} \label{sec:results}
We now use a surface hopping algorithm to solve the CME dynamics. In the surface hopping method, we use a swarm of trajectories to represent nuclear densities.  We evolve trajectories on the active potential energy surfaces (PESs), and we introduce hopping between the PESs. To do so, at each time step, we generate a random number from 0 to 1 from uniform distribution; we then compare the random number to hopping rate times $dt$ ($dt$ is the time interval). 
A hopping to a different potential energy surface occurs when the random number is smaller than the hopping rate times $dt$; otherwise, the trajectory remains on the original PES. 

In a sliding metal-semiconductor junction, the mechanic motion could induce phonon excitation, especially at the local contact asperities,  which have a typical size of 10-100 nm. 
The single asperities at contact interfaces experience a large local pressure at GPa level and a local contact temperature $T$ as high as 1000K.\cite{kennedy2001frictional} 
Such excited phonon motion could then introduce electron-hole pairs as well as voltage generation through electron-phonon couplings. 
To mimic a such process, we initialize all trajectories on the ground state (one electron in the valance band,  with PES $U_1= U_0 + E_v$), where the momentums and positions of these trajectories satisfy a 
Boltzmann distribution at a given initial phonon temperature $T_i $, 
\begin{eqnarray}
  \rho_v (x, p) &=& \frac{\omega}{2\pi kT_i } \exp \Big(-\frac{1}{kT_i} \big( \frac{p^2}{2m}   +  \frac12 m\omega^2 (x-x_0)^2 \big) \Big) \\
  \rho_0 &=& \rho_c = \rho_{vc} = 0 
\end{eqnarray}
Here, $x_0= -\sqrt{2}g/\hbar\omega$ is the position of lowest energy on the ground state. We then evolve our equation of motion to see how phonon energy relaxes and how electron-hole pairs are generated as a function of time. In our simulation, we use 4th order Runge-Kutta to propagate the dynamics. Unless otherwise stated, we use 10000 trajectories for our SH simulations.

\subsection{Phonon Relaxation}
\label{sec:Phonon Relaxation}

\begin{figure}[h]
   \centering{}
   \subfloat[]{\includegraphics[width=0.5\textwidth]{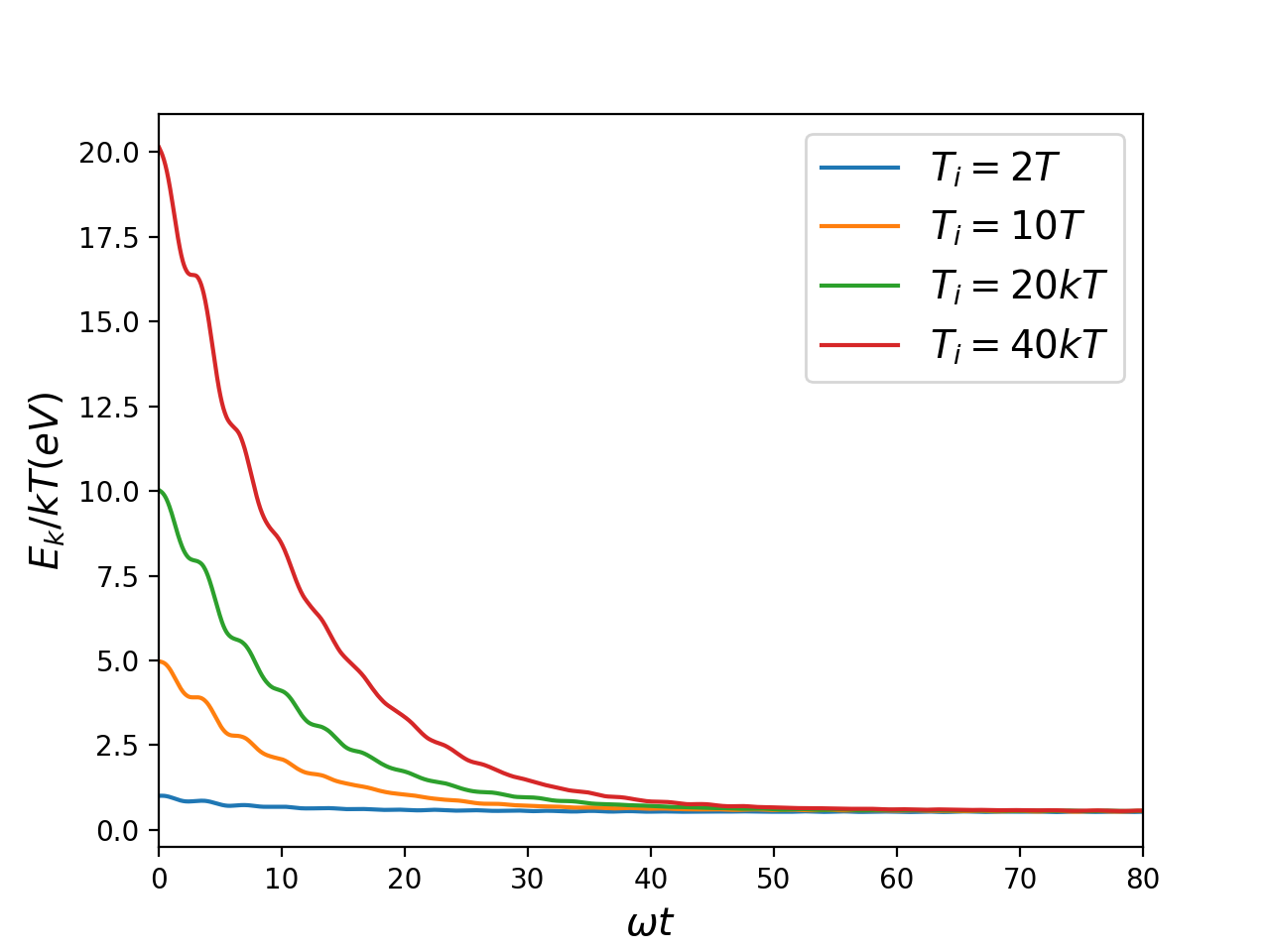}}
   \hfill
   \subfloat[]{\includegraphics[width=0.5\textwidth]{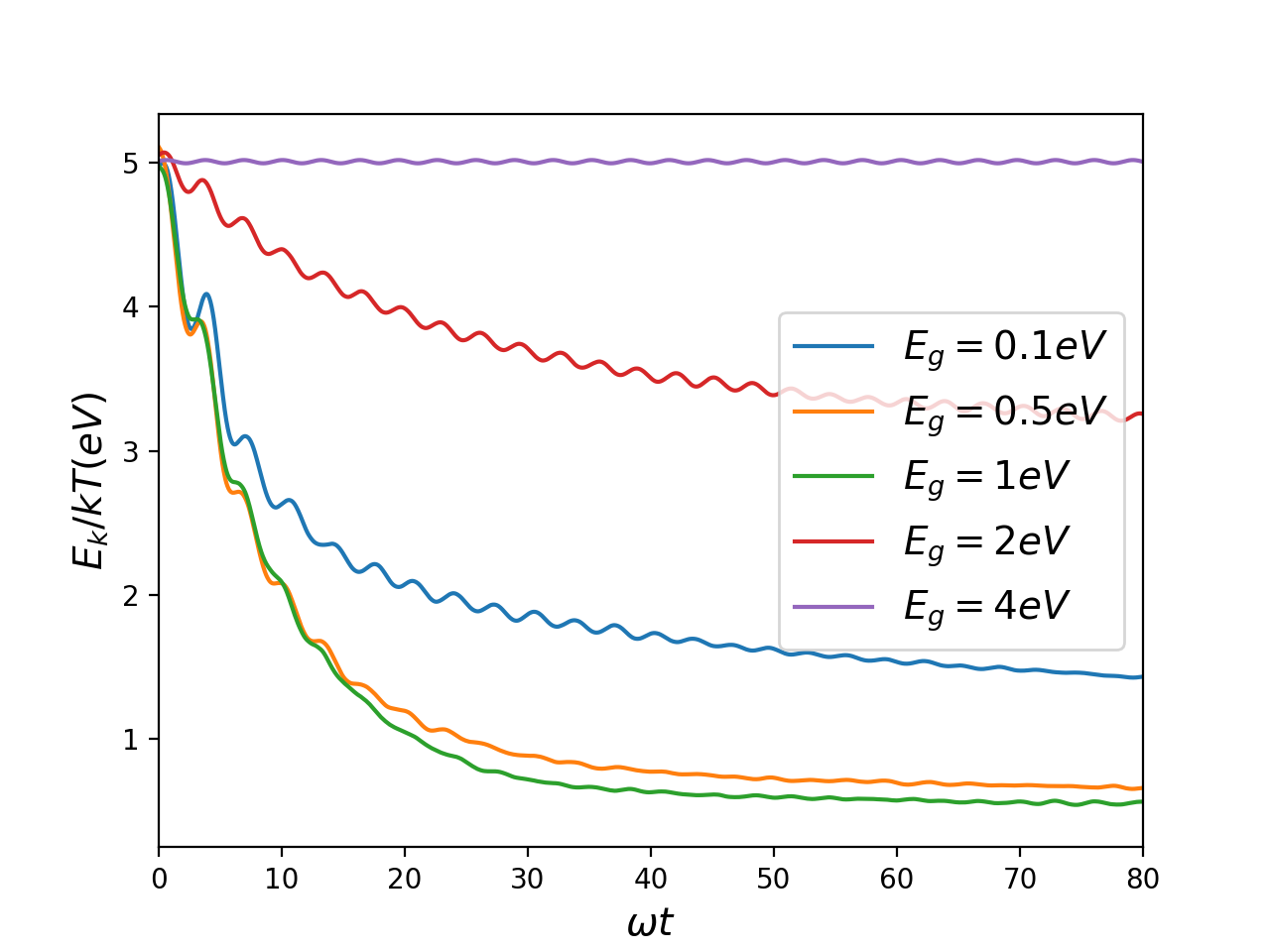}}
   \hfill
   \caption{Phonon relaxation:$\hbar\omega=0.03\ eV, g= 0.09\ eV, kT = 0.05\ eV$, $\Gamma_c = \Gamma_v = 0.02 \ eV$. (a)Phonon Relaxation with different initial conditions, $T_i$ represents 
   the temperature at which the oscillator is initialized. (b)Phonon relaxation with different Bandgaps, $E_g(eV)$. }
   \label{fig:phonon}
\end{figure}

We first look at the phonon relaxation at the interfaces. In Figure~\ref*{fig:phonon}(a), we initialize our phase space density on the ground state with different phonon temperatures $T_i$, and we plot average kinetic energy as a function of time. Here, we fix the temperature of the metal (electron bath, $kT=0.05eV$), which only appears in the equation of motion through the Fermi distribution in the hopping rates.  we note that, the kinetic energy of the phonon will reach to the same steady state, with the average kinetic energy being half $kT$, i.e. the temperature of the metal, regardless of the initial conditions.  In fact, we can verify that, the steady state nuclear distribution will obey a Boltzmann distribution with the same temperature of the metal 
(see Ref. \citen{dou2015surface,dou2015surface3}). Physically, this observation indicates that when the phonon degrees of freedom interact with a manifold of  electronic states long enough, the phonon will reach to an equilibrium distribution with the same temperature as the electron bath. Note also that, the relaxation rate is independent of the initial kinetic energy. As shown in Ref. \citen{ouyang2016dynamics,dou2015frictional}, the relaxation rate strongly depends on the electron-phonon coupling $g$, the timescales of electronic motion $\Gamma_c$ and/or $\Gamma_v$, and the timescale of the nuclear motion $\hbar\omega$.

In Figure~\ref*{fig:phonon}(b), we plot the phonon relaxation for different band gaps $E_g$. We have fixed the initial phonon temperature, $T_i  =10 T$. Here, we observe a turnover of the relaxation timescales for different Bandgaps. 1) For the case where band gaps $E_g>1eV$, the relaxation takes longer for larger bandgaps. In fact, for the Bandgap as large as $E_g=4eV$, we do not see energy relaxation at all. 
This is because when the band gap is large enough, the crossing between PESs will only occur at very large potentials (see Fig. \ref{fig:PES1}), such that hopping events are very rare, which results in little or no energy relaxation. 
2) For the cases where band gaps $E_g < 1eV$, the relaxation timescales could also be larger for smaller bandgaps. This is due to the fact that when hopping occurs, the energy loss is small if the bandgaps are small. The relaxation timescale is a result of two competing effects: the frequency of hopping and energy loss when hopping events occur. Increasing the bandgap will reduce the frequency of hopping, but increase energy loss when hopping events occur.

\begin{figure}[H]
   \centering
   \subfloat[]{\includegraphics[width=8cm]{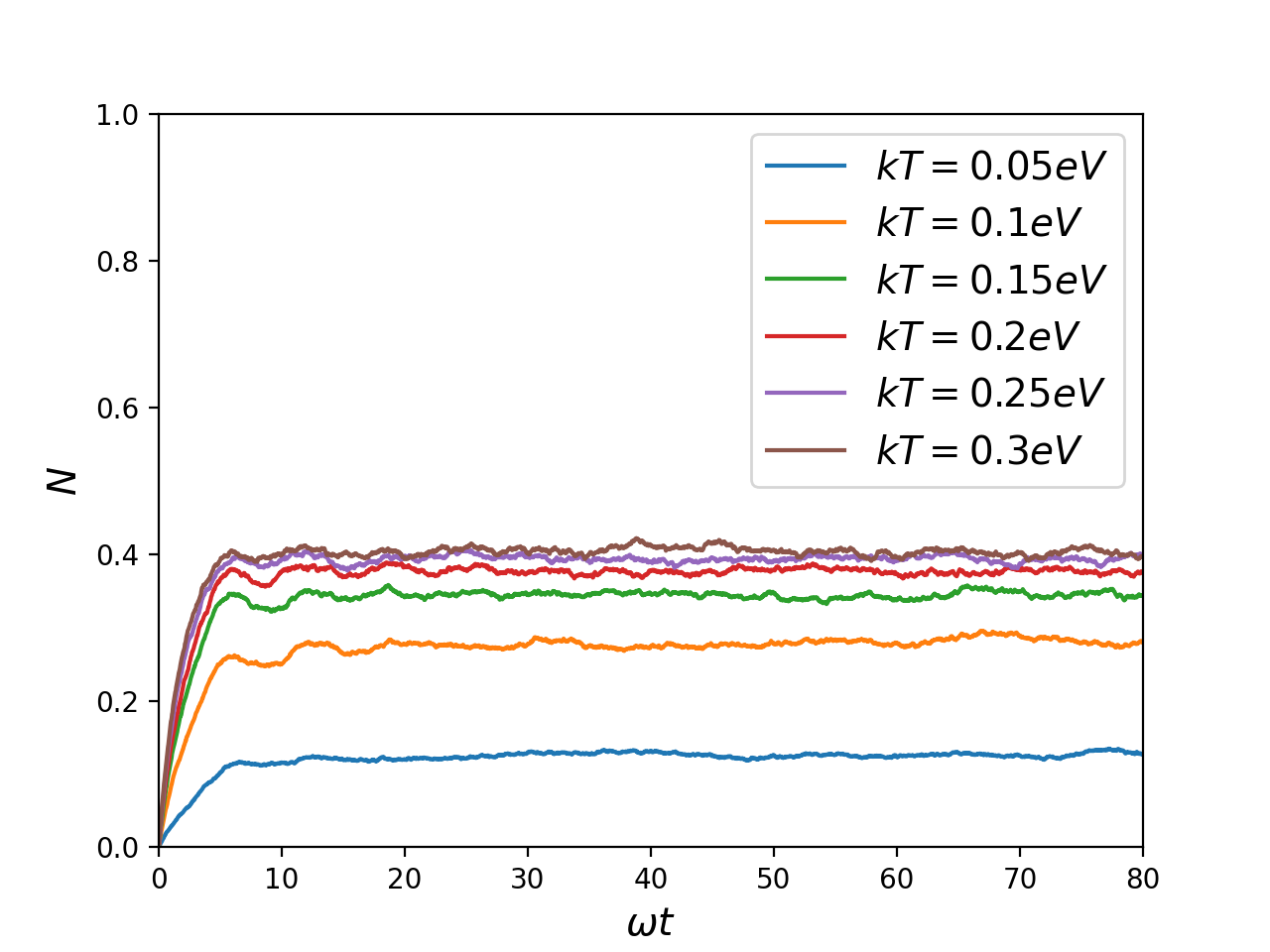}}
   \hfill
   \subfloat[]{\includegraphics[width=8cm]{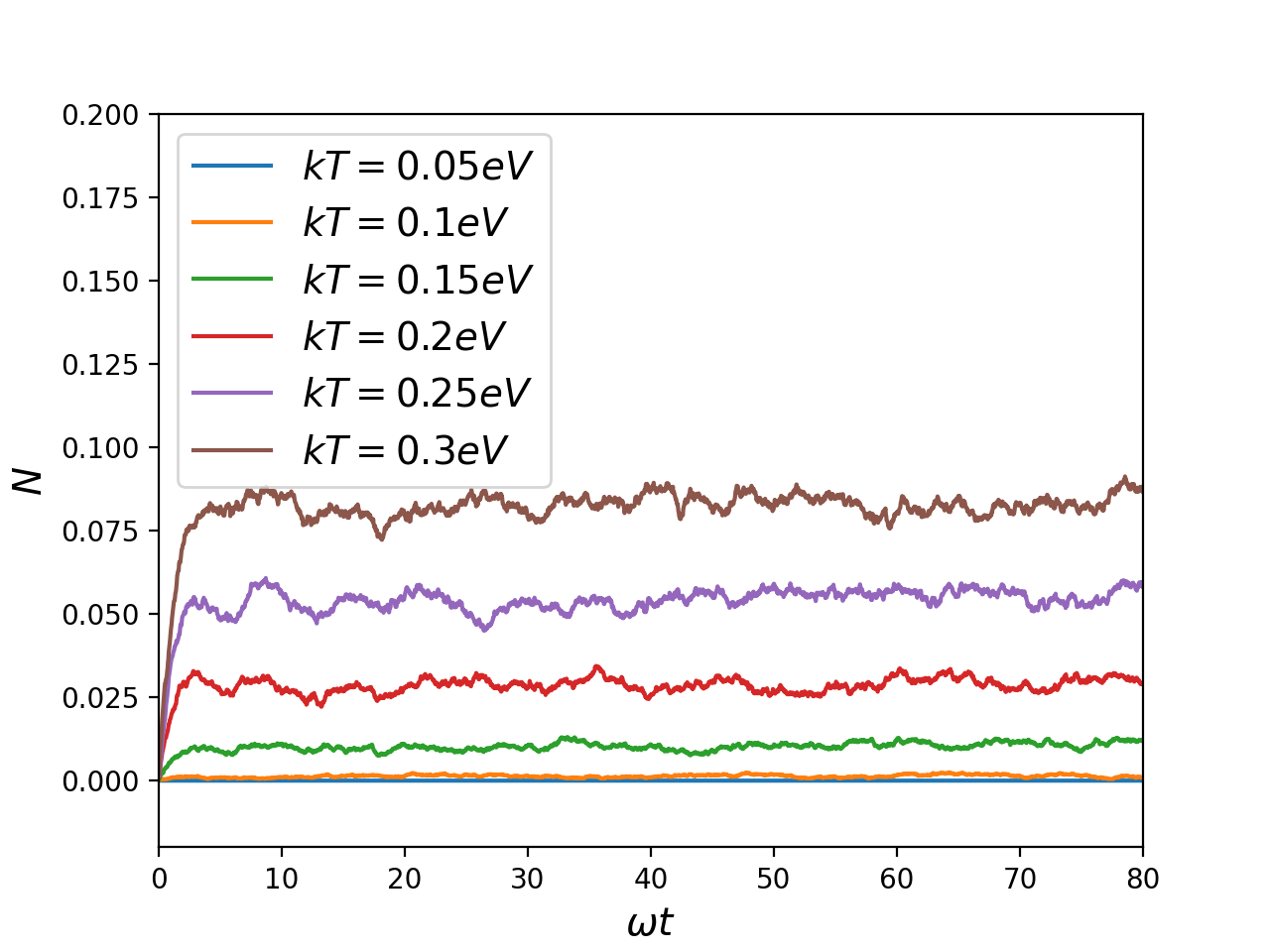}}
   \hfill
   \subfloat[]{\includegraphics[width=8cm]{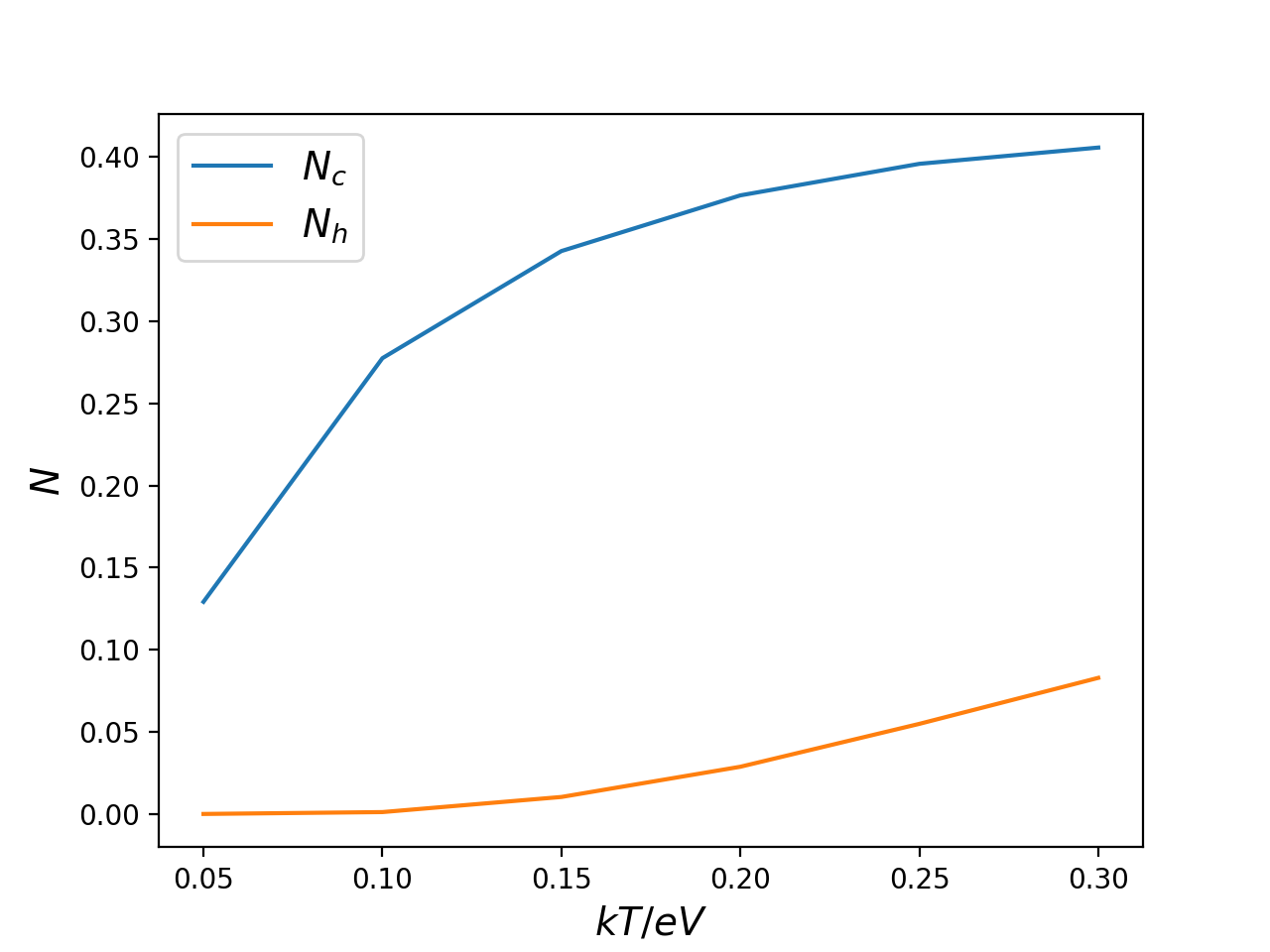}}
   \caption{Electron and hole populations at different local temperatures.  (a)Electron population as a function of time. 
   (b)Hole population as a function of time. (c)Steady state population of electrons and holes as a function of local temperatures. Parameters: $\hbar\omega=0.03\ eV, g=0.09\ eV, E_g=1\ eV$, $\Gamma_c = \Gamma_v = 0.02 \ eV$.}
   \label{fig:popNkT}
\end{figure}

\subsection{Electron and hole populations}
In a sliding metal-semiconductor interfaces, the open-circuit voltage ($V_{oc}$) generation (tribovoltaic potential difference between the metal and the semiconductor) 
is associated with the electron-hole (e-h) pair generation in a mechanically excited system.\cite{liu2019interfacial,zhang2020tribovoltaic}
To quantify the electron-hole (e-h) pair generation, we define the electron population on the conduction band $N_c$ as well as the hole population on the valence band $N_h$ as following: 
\begin{eqnarray}
   N_c = \int \big(\rho_c (x,p) + \rho_{vc} (x,p) \big) dxdp, \label{eq:16} \\ 
   N_h = \int \big (\rho_0(x,p) + \rho_c(x,p) \big) dxdp
   \label{eq:17}
\end{eqnarray}
Here again, $\rho_c(x,p)$ is the phase space density for the one electron and one hole state, $\rho_{vc} (x,p)$ is the density for two electron states,  and $\rho_0(x,p)$ is the density for one  hole state (see Figure~\ref{fig:PES1}).

In Figure~\ref*{fig:popNkT}(a) and Figure~\ref*{fig:popNkT}(b), we plot hole and electron populations at different temperatures as a function of time. Here the initial phonon temperature $T_i$ is set to be equal to the temperature of the electron or the local temperature of the asperities $T$. 
We further vary the contact temperature at the single asperities depending on mechanical motion parameters (e.g. pressure and speed) at a sliding metal-semiconductor junction. We find that the timescales for electron and hole generation are roughly the same. In general, the steady state electron generation is larger than the hole generation. 
When the local temperature is small enough($\leq 0.1 eV$), there will be nearly no hole generation. This can be seen from Figure~\ref{fig:PES1}(a): $U_3$ is the lowest excited states, whereas $U_0$ and $U_2$ are excited states with higher energies. At small local temperatures, there will be little distribution on potential surface $U_0$ and $U_2$, and most of the excited states is on potential surface $U_3$. 

To further illustrate the temperature effects, 
in Figure~\ref*{fig:popNkT}(c) , we further plot electron and hole populations as a function of  temperature at the steady states. Note that both electron and hole populations increase
with temperature. This observation is in agreement with experimental results. With faster mechanical motion and higher pressure at the interfaces, the temperature of local asperities is larger, such that more electron-hole pairs are generated at the interfaces. Again, we see the asymmetry of electron and hole generations at given temperature. We hope future experimental results can justify our simulation.

\begin{figure}[H]
   \centering
   \subfloat[]{\includegraphics[width=0.5\linewidth]{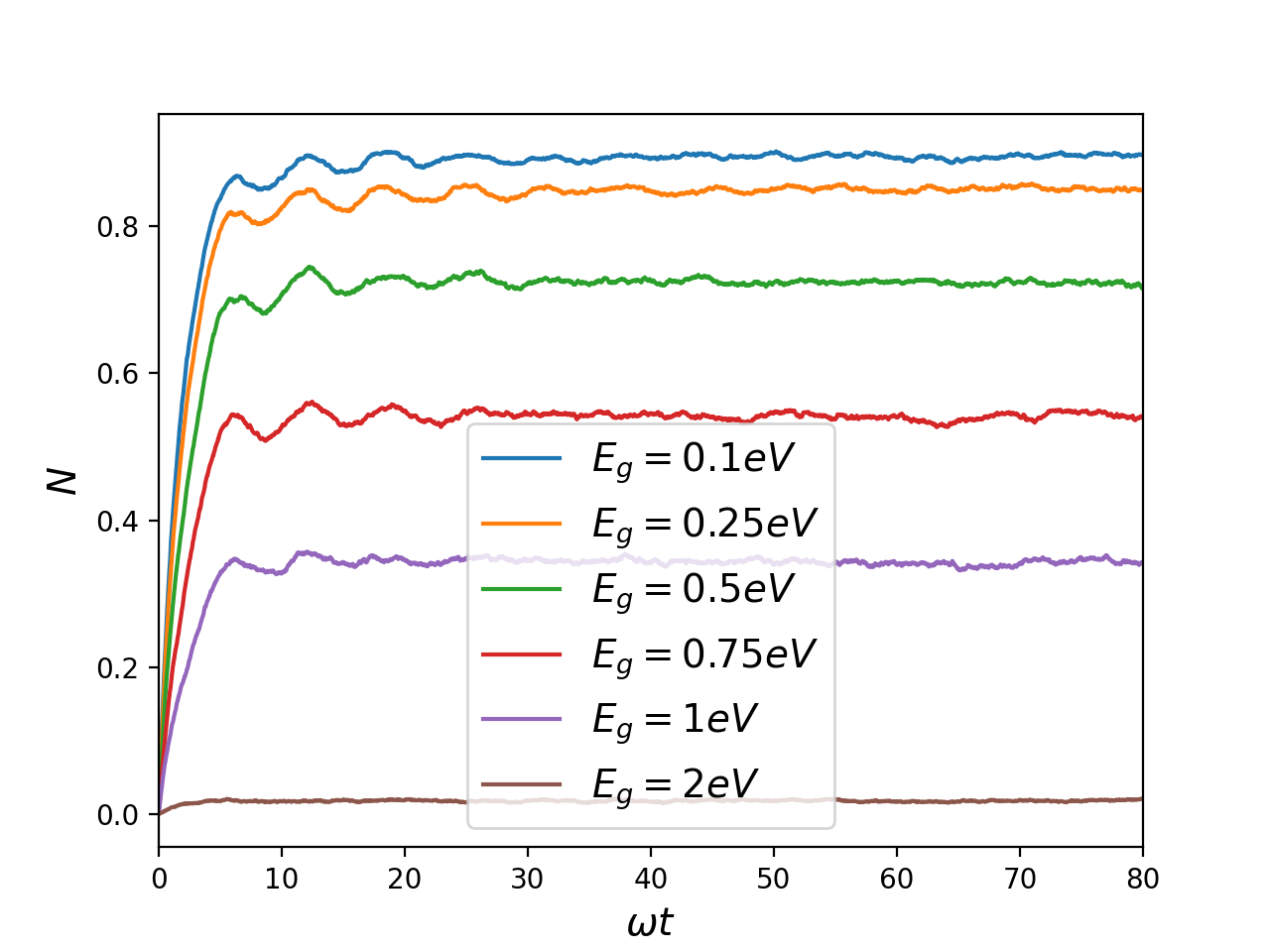}}
   \hfill
   \subfloat[]{\includegraphics[width=0.5\linewidth]{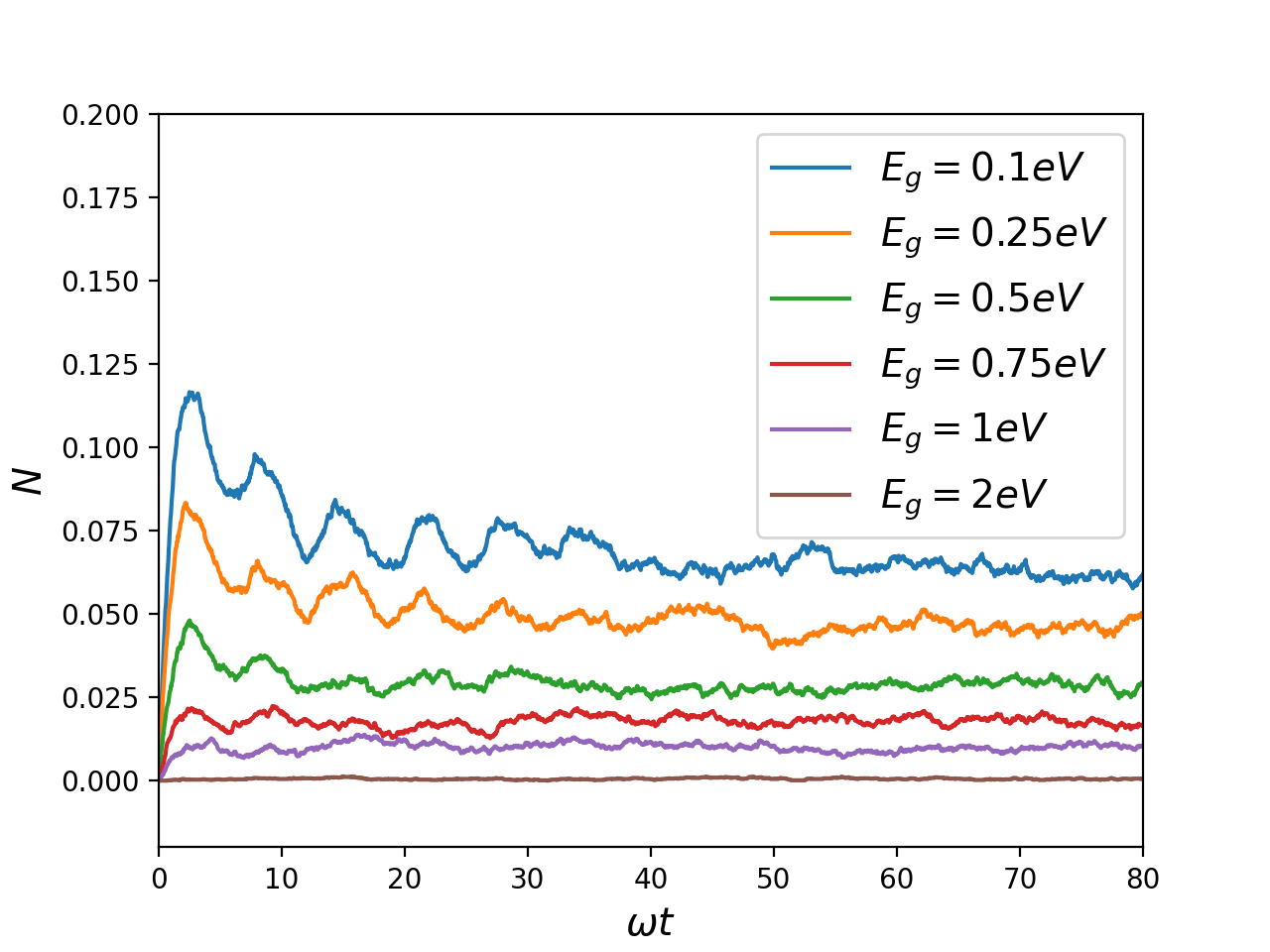}}
   \hfill
   \subfloat[]{\includegraphics[width=0.5\linewidth]{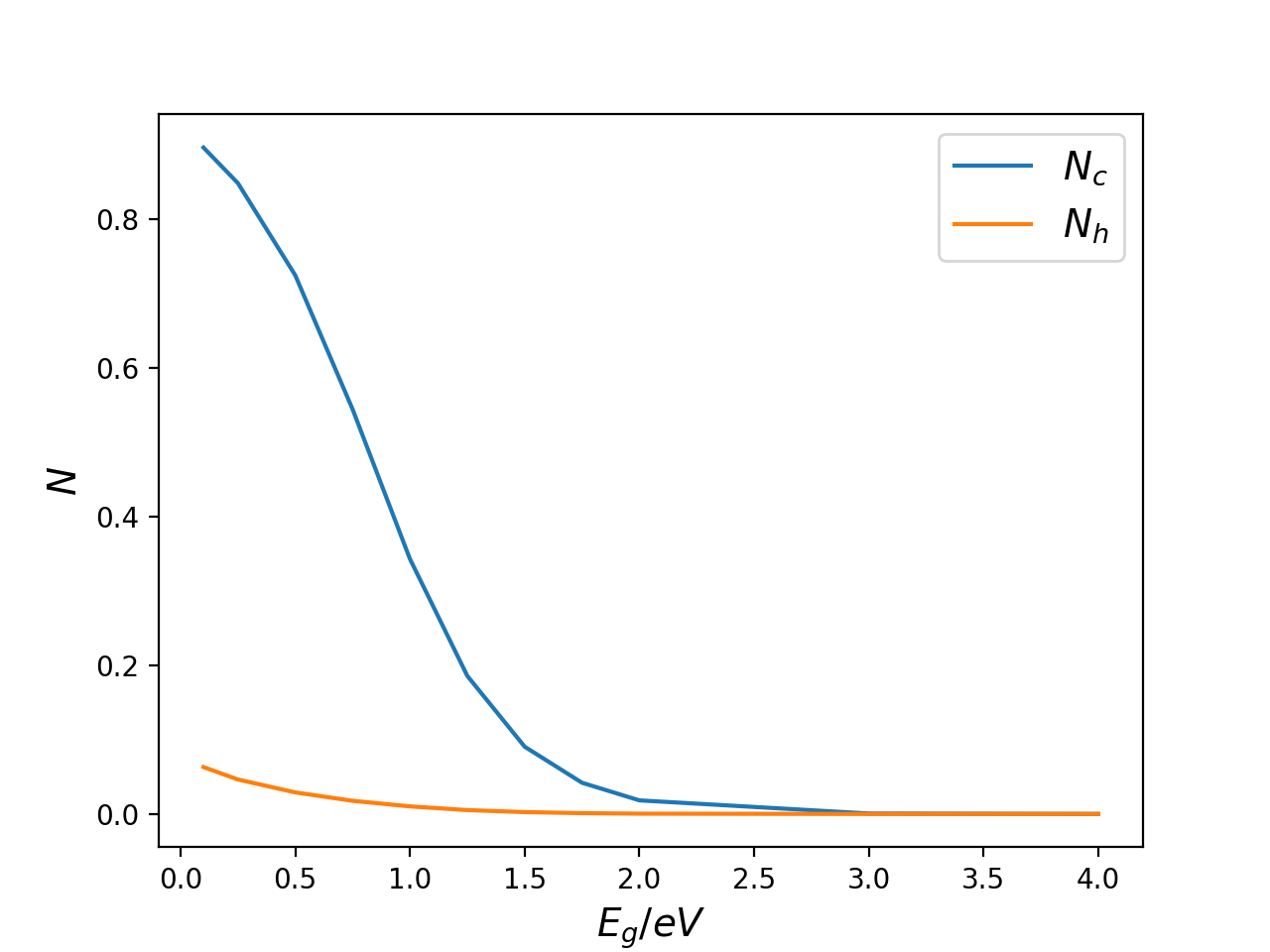}}
   \caption{Electron and hole populations for different bandgaps ($eV$). (a)Electron population as a function of time. (b) Hole population as a function of time. (c) Steady state electron and hole populations as a function of bandgap. Parameters: $\hbar\omega=0.03, g=0.09, kT = 0.15eV $.}
   \label{fig:popNEg}
\end{figure}

We now study the effects of bandgaps on the electron and hole generations. In Figure~\ref*{fig:popNEg}(a) and Figure~\ref*{fig:popNEg}(b), 
we plot hole and electron populations  at different bandgaps as a function of time. 
Notice that the timescales for electron and hole generations strongly depends on bandgaps. As indicated before (see Section~\ref{sec:Phonon Relaxation}), the bandgap strongly affects the hopping rates, which in turn determines the timesacle for population relaxation. 
In Figure~\ref*{fig:popNEg}(c), we plot the steady state electron and hole populations as a function of bandgap.  We see that the steady state electron and hole population decrease with the bandgaps. Obviously, with larger bandgaps, the chance of occupying excited state decreases, such that there will be less electron and hole generations. In fact, from our calculation, we see that there is almost no electron and hole generation for bandgap as large as 3eV. Such a fact will show their signature in the voltage generation as depicted below. 


\subsection{Voltage generation}
With the definition of the electron and hole generations in Eq. \ref{eq:16} and \ref{eq:17}, it is natural to define the open-circuit voltage ($V_{oc}$) as 
\begin{eqnarray}
   V_{oc} = \frac{1}{2}E_g(N_c + N_h)
   \label{eq:19}
\end{eqnarray}
Here again, $E_g$ is the bandgap. The factor $1/2$ is introduced due to the fact that we have set Fermi level in the middle of the conduction level and valence level. Obviously, the voltage depends on the bandgap as well as the electron and hole generations.

\begin{figure}[H]
   \centering
   \subfloat[]{\includegraphics[width=0.5\linewidth]{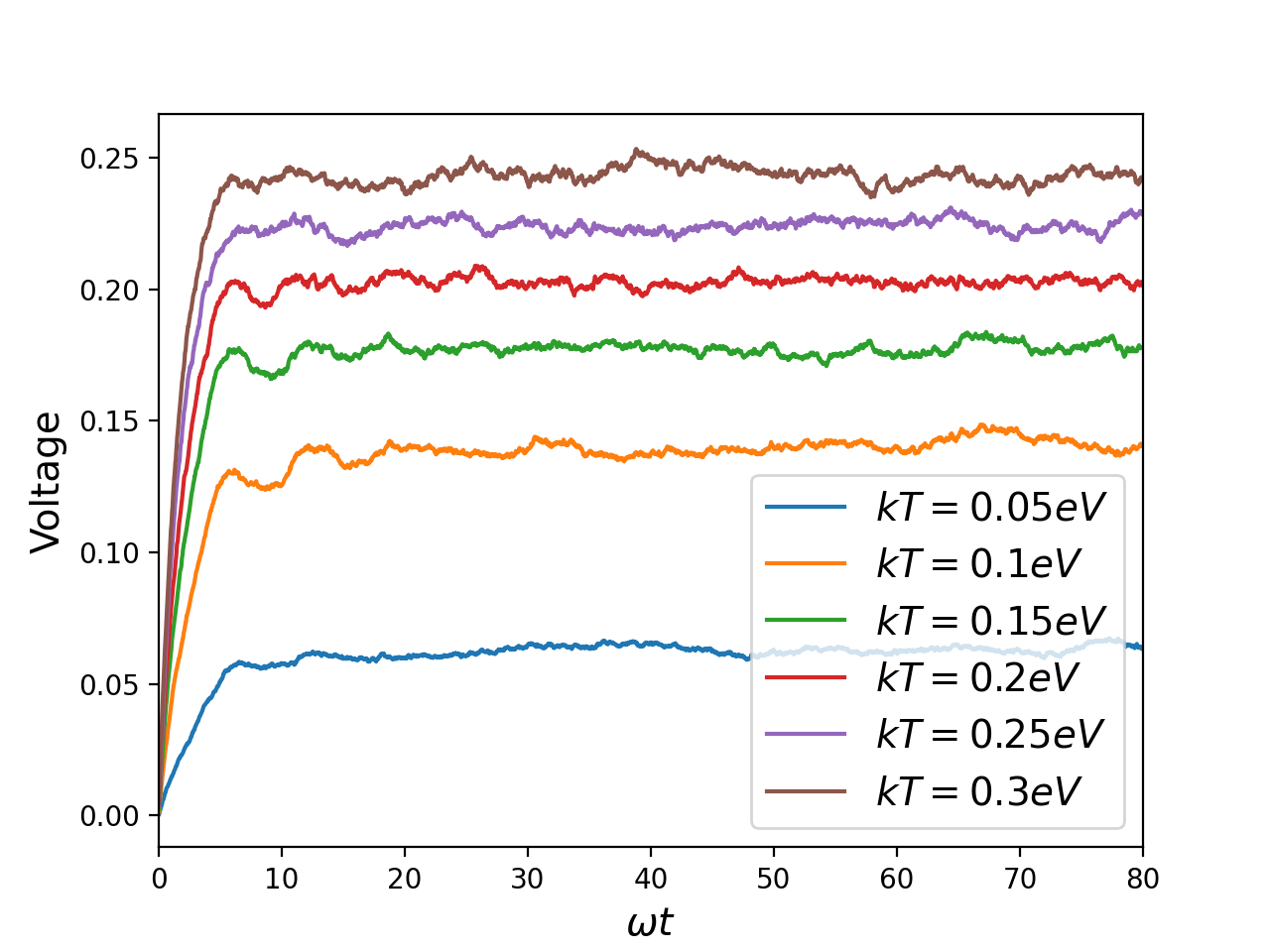}}
   \hfill
   \subfloat[]{\includegraphics[width=0.5\linewidth]{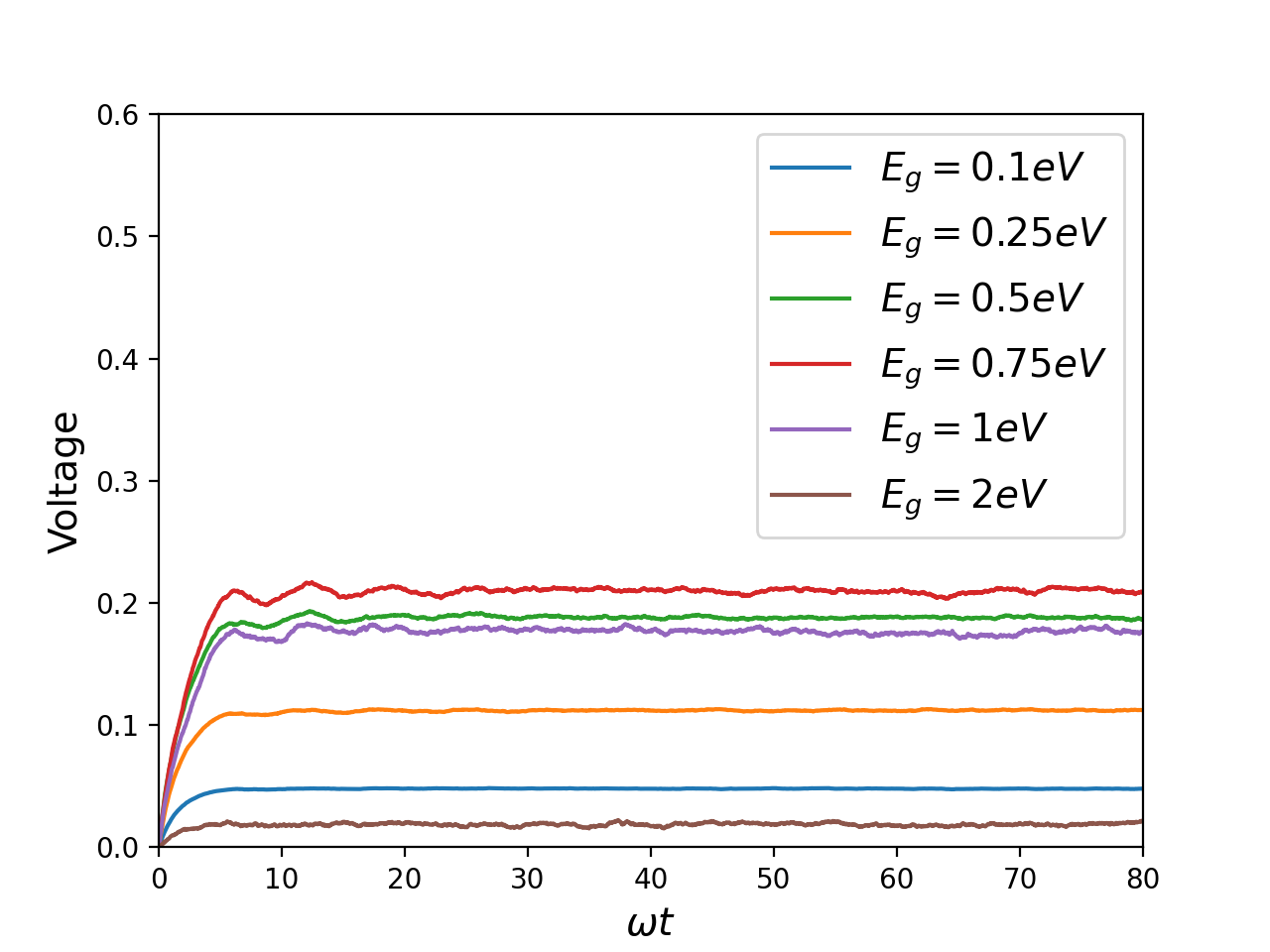}}
   \hfill
   \subfloat[]{\includegraphics[width=0.5\linewidth]{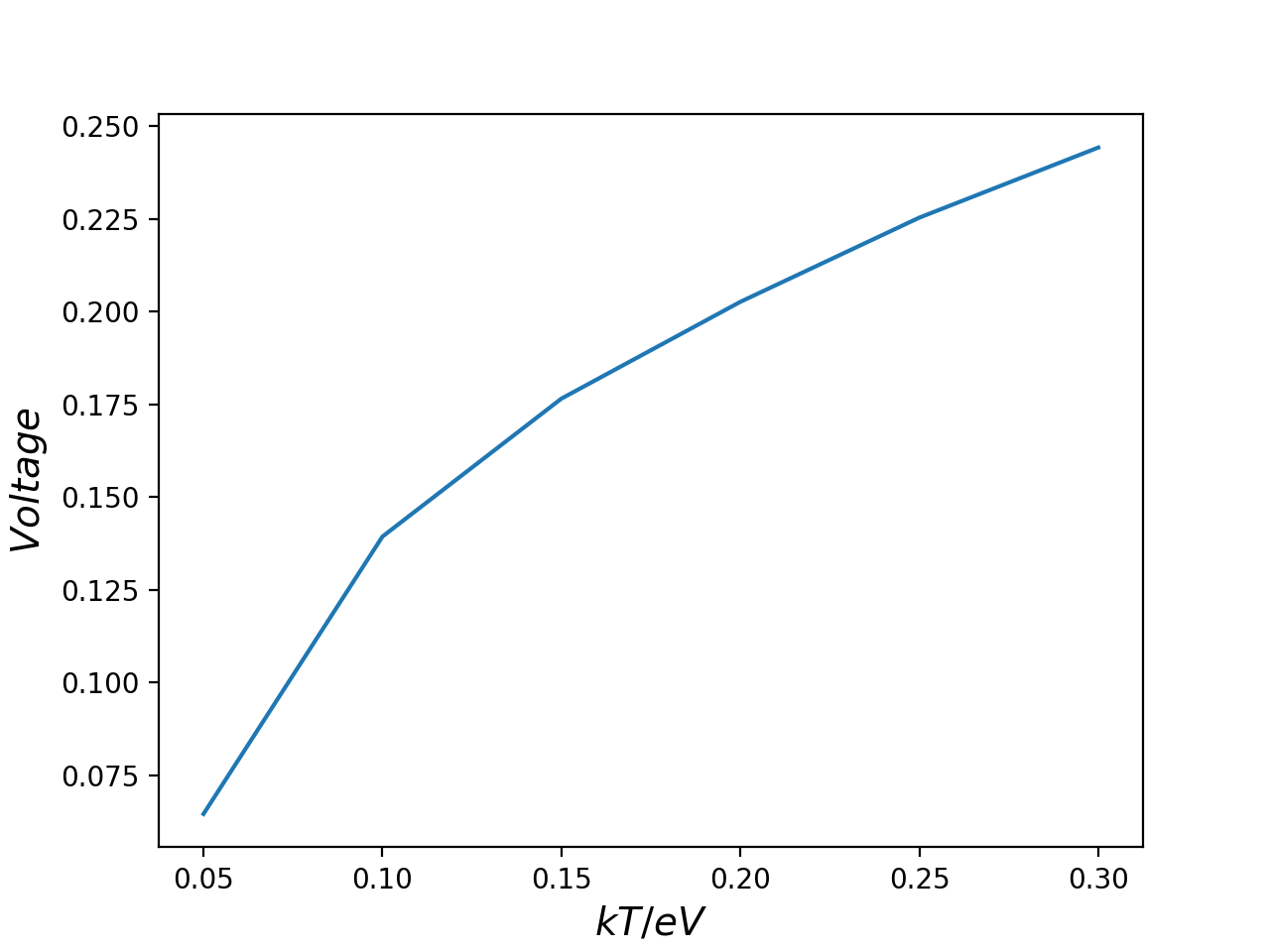}}
   \hfill
   \subfloat[]{\includegraphics[width=0.5\linewidth]{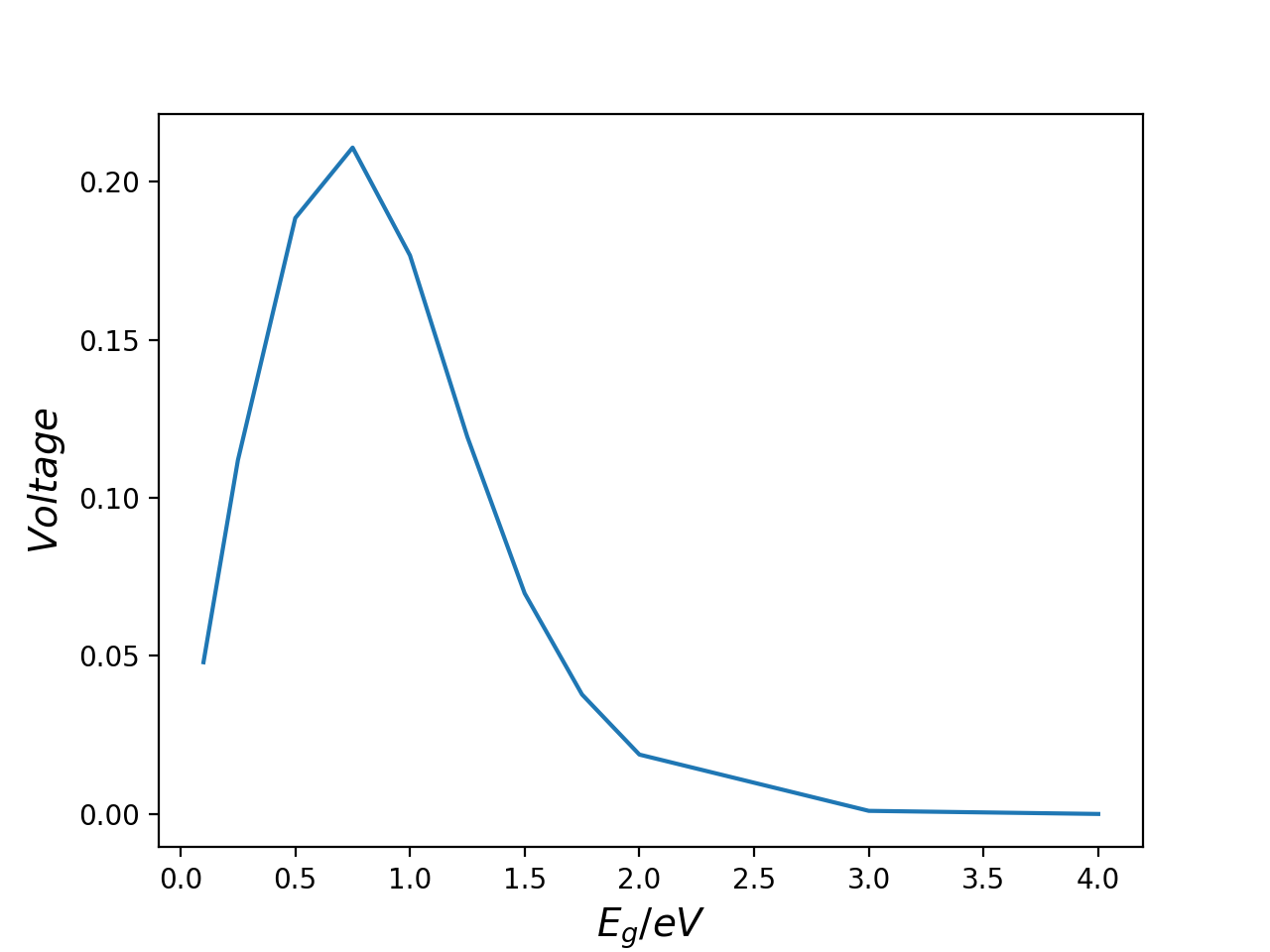}}
   \caption{Average voltage ($V_{oc}$) for different local temperatures and bandgaps. (a) $V_{oc}$ as a function of time for different temperatures, $E_g =1eV$.
   (b)$V_{oc}$ as a function of time for different bandgaps, $kT=0.15eV$. (c) Steady state $V_{oc}$ as a function of temperature, $E_g =1eV$. (d) Steady state $V_{oc}$  as a function of bandgap, $kT=0.15eV$. Other parameters: $\hbar\omega=0.03 \ eV, g=0.09 \ eV $. }
   \label{fig:potential}
\end{figure}

In Figure~\ref*{fig:potential}(a), we plot voltage as a function of time for different temperatures. Note that at longer time, the voltages reach to stationary values. To further illustrate the effects of the temperature, we plot the steady state voltages as a function of the temperature in 
Figure~\ref*{fig:potential}(c). Here we see that the steady state voltage increases with the temperature. This result is consistent with the observation in Figure~\ref{fig:popNkT}(c), where both $N_c$ and $N_h$ increase with temperature. In general, larger local temperature will lead to higher voltage generation. Again, this is in agreement with the experimental results.

In Figure~\ref*{fig:potential}(b), we plot $V_{oc}$  as a function of time for different bandgaps ($E_g$). Notice that for smaller bandgaps (< 0.75 eV), the steady state voltage increase with the bandgap; Whereas for larger bandgaps (> 0.75 eV) the steady state voltage decreases with bandgap. This is further shown in Figure~\ref*{fig:potential}(d), 
where we clearly see a turnover for the voltage as a function of the bandgap. The turning point is around 0.75 eV for the bandgap. This turnover is a result of two competing factors: the electron-hole populations as well as the bandgaps. See the definition of voltage in Eq. \ref{eq:19}.  
With small bandgap, there is little voltage generation. Increasing the bandgap will lead to less electron and hole generations. When the bandgap is large enough, there will be no electron and hole generations, hence no voltage generation. Such a turnover is being observed in experiments. We hope to quantify this turnover more precisely for realistic system in the future study.

\section{Conclusions} \label{sec:con}
We have offered a quantum mechanic perspective of the tribovaltaic effects at a sliding metal-semiconductor interfaces. We have introduced a two-level Anderson-Holstein model to describe nonadiabatic electron and energy transfer at the interfaces. Furthermore,  
we have used a classical master equation (CME) as well as a surface hopping algorithm to model the coupled electron-phonon dynamics. Using such a dynamical method, we can quantify the electron-hole pair generations as well as voltage generations. We find that, the electron and hole populations increase with local temperature, but decrease with the bandgaps. However, there is a turnover for the voltage generation as a function of the bandgaps. This finding is strongly supported by the experimental results. We believe this work present an atomic description of tribovaltaic. Future work must apply our approach to realistic systems for better design and control of sliding metal-semiconduction junction.

\begin{acknowledgement}
WD acknowledges the startup funding from Westlake University. 

\end{acknowledgement}

\section*{Data availability} 
The data that support the findings of this study are available
from the corresponding author upon reasonable request.

\bibliography{e-ph_coupling}

\end{document}